\documentclass[12pt,preprint]{emulateapj}

\newcommand{\swift}{{\it Swift}}
\newcommand{\hst}{{\it HST}}
\newcommand{\ngc}{NGC\, 5548}
\newcommand{\et}{et al.\,}
\newcommand{\rb}{}

\shorttitle{AGN STORM. II. \swift\ Observations}
\shortauthors{Edelson \et}

\begin{document}

\title{Space Telescope and Optical Reverberation Mapping Project.\\
II.\ {\it Swift} and \hst\ reverberation mapping of the accretion disk of \ngc}

\author{
R.~Edelson\altaffilmark{1},
J.~M.~Gelbord\altaffilmark{2,3},
K.~Horne\altaffilmark{4},
I.~M.~M$^{\rm c}$Hardy\altaffilmark{5},
B.~M.~Peterson\altaffilmark{6,7},
P.~Ar\'{e}valo\altaffilmark{8},
A.~A.~Breeveld\altaffilmark{9}
G.~De~Rosa\altaffilmark{6,7,10},
P.~A.~Evans\altaffilmark{11},
M.~R.~Goad\altaffilmark{11},
G.~A.~Kriss\altaffilmark{10,12},
W.~N.~Brandt\altaffilmark{13},
N.~Gehrels\altaffilmark{14},
D.~Grupe\altaffilmark{15},
J.~A.~Kennea\altaffilmark{13},
C.~S.~Kochanek\altaffilmark{6,7},
J.~A.~Nousek\altaffilmark{13},
I.~Papadakis\altaffilmark{16,17},
M.~Siegel\altaffilmark{13},
D.~Starkey\altaffilmark{4},
P.~Uttley\altaffilmark{18},
S.~Vaughan\altaffilmark{11},
S.~Young\altaffilmark{1},
A.~J.~Barth\altaffilmark{19},
M.~C.~Bentz\altaffilmark{20},
B.~J.~Brewer\altaffilmark{21},
D.~M.~Crenshaw\altaffilmark{20},
E.~Dalla~Bont\`{a}\altaffilmark{22,23},
A.~De~Lorenzo-C\'{a}ceres\altaffilmark{4},
K.~D.~Denney\altaffilmark{6,7,24},
M.~Dietrich\altaffilmark{25,26},
J.~Ely\altaffilmark{10},
M.~M.~Fausnaugh\altaffilmark{6},
C.~J.~Grier\altaffilmark{6,13},
P.~B.~Hall\altaffilmark{27},
J.~Kaastra\altaffilmark{28,29,30},
B.~C.~Kelly\altaffilmark{31},
K.~T.~Korista\altaffilmark{32},
P.~Lira\altaffilmark{33},
S.~Mathur\altaffilmark{6,7},
H.~Netzer\altaffilmark{34},
A.~Pancoast\altaffilmark{31},
L.~Pei\altaffilmark{20},
R.~W.~Pogge\altaffilmark{6,7},
J.~S.~Schimoia\altaffilmark{6,35},
T.~Treu\altaffilmark{31,36,37},
M.~Vestergaard\altaffilmark{38,39},
C.~Villforth\altaffilmark{4},
H.~Yan\altaffilmark{40},
and Y.~Zu\altaffilmark{6,41}
}

\altaffiltext	{1}{Department of Astronomy, University of Maryland, College Park, MD 20742-2421}
\altaffiltext	{2}{Spectral Sciences Inc., 4 Fourth Ave., Burlington, MA 01803}
\altaffiltext	{3}{Eureka Scientific Inc., 2452 Delmer St. Suite 100, Oakland, CA 94602}
\altaffiltext	{4}{SUPA Physics and Astronomy, University of St. Andrews, Fife, KY16 9SS Scotland, UK}
\altaffiltext	{5}{University of Southampton, Highfield, Southampton, SO17 1BJ, UK}
\altaffiltext	{6}{Department of Astronomy, The Ohio State University, 140 W 18th Ave, Columbus, OH 43210}
\altaffiltext	{7}{Center for Cosmology and AstroParticle Physics, The Ohio State University, 191 West Woodruff Ave, Columbus, OH 43210}
\altaffiltext	{8}{Instituto de F\'{\i}sica y Astronom\'{\i}a, Facultad de Ciencias, Universidad de Valpara\'{\i}so, Gran Bretana N 1111, Playa Ancha, Valpara\'{\i}ıso, Chile}
\altaffiltext	{9}{Mullard Space Science Laboratory, University College London, Holmbury St. Mary, Dorking, Surrey RH5 6NT, UK}
\altaffiltext	{10}{Space Telescope Science Institute, 3700 San Martin Drive, Baltimore, MD 21218}
\altaffiltext {11}{University of Leicester, Department of Physics and Astronomy, Leicester, LE1 7RH, UK}
\altaffiltext	{12}{Department of Physics and Astronomy, The Johns Hopkins University, Baltimore, MD 21218}
\altaffiltext	{13}{Department of Astronomy and Astrophysics, Eberly College of Science, Penn State University, 525 Davey Laboratory, University Park, PA 16802}
\altaffiltext	{14}{Astrophysics Science Division, NASA Goddard Space Flight Center, Greenbelt, MD 20771}
\altaffiltext	{15}{Space Science Center, Morehead State University, 235 Martindale Dr., Morehead, KY 40351}
\altaffiltext	{16}{Department of Physics and Institute of Theoretical and Computational Physics, University of Crete, GR-71003 Heraklion, Greece}
\altaffiltext	{17}{IESL, Foundation for Research and Technology, GR-71110 Heraklion, Greece}
\altaffiltext	{18}{Astronomical Institute `Anton Pannekoek,' University of Amsterdam, Postbus 94249, NL-1090 GE Amsterdam, The Netherlands}
\altaffiltext	{19}{Department of Physics and Astronomy, 4129 Frederick Reines Hall, University of California, Irvine, CA 92697}
\altaffiltext	{20}{Department of Physics and Astronomy, Georgia State University, 25 Park Place, Suite 605, Atlanta, GA 30303}
\altaffiltext	{21}{Department of Statistics, The University of Auckland, Private Bag 92019, Auckland 1142, New Zealand}
\altaffiltext	{22}{Dipartimento di Fisica e Astronomia ``G. Galilei,'' Universit\`{a} di Padova, Vicolo dell'Osservatorio 3, I-35122 Padova, Italy}
\altaffiltext	{23}{INAF-Osservatorio Astronomico di Padova, Vicolo dell'Osservatorio 5 I-35122, Padova, Italy}
\altaffiltext	{25}{Department of Physics and Astronomy, Ohio University, Athens, OH 45701}
\altaffiltext	{26}{Department of Physical and Earth Sciences, Worcester State University, 486 Chandler Street, Worcester, MA 01602}
\altaffiltext	{27}{Department of Physics and Astronomy, York University, Toronto, ON M3J 1P3, Canada}
\altaffiltext	{28}{SRON Netherlands Institute for Space Research, Sorbonnelaan 2, 3584 CA Utrecht, The Netherlands}
\altaffiltext	{29}{Department of Physics and Astronomy, Univeristeit Utrecht, P.O. Box 80000, 3508 Utrecht, The Netherlands}
\altaffiltext	{30}{Leiden Observatory, Leiden University, PO Box 9513, 2300 RA Leiden, The Netherlands}
\altaffiltext	{31}{Department of Physics, University of California, Santa Barbara, CA 93106}
\altaffiltext	{32}{Department of Physics, Western Michigan University, 1120 Everett Tower, Kalamazoo, MI 49008-5252}
\altaffiltext	{33}{Departamento de Astronomia, Universidad de Chile, Camino del Observatorio 1515, Santiago, Chile}
\altaffiltext	{34}{School of Physics and Astronomy, Raymond and Beverly Sackler Faculty of Exact Sciences, Tel Aviv University, Tel Aviv 69978, Israel}
\altaffiltext	{35}{Instituto de F\'{\i}sica, Universidade Federal do Rio do Sul, Campus do Vale, Porto Alegre, Brazil}
\altaffiltext	{36}{Department of Physics and Astronomy, University of California, Los Angeles, CA 90095-1547}
\altaffiltext	{38}{Dark Cosmology Centre, Niels Bohr Institute, University of Copenhagen, Juliane Maries Vej 30, DK-2100 Copenhagen, Denmark}
\altaffiltext	{39}{Steward Observatory, University of Arizona, 933 North Cherry Avenue, Tucson, AZ 85721}
\altaffiltext	{40}{Department of Physics and Astronomy, University of Missouri, Columbia, MO 65211}
\altaffiltext	{41}{Department of Physics, Carnegie Mellon University, 5000 Forbes Avenue, Pittsburgh, PA 15213}

\footnotetext[24]{NSF Postdoctoral Research Fellow}
\footnotetext[37]{Packard Fellow}

\begin{abstract}
Recent intensive \swift\ monitoring of the Seyfert 1 galaxy NGC 5548 yielded 282 usable epochs over 125 days across six UV/optical bands and the X-rays. 
This is the densest extended AGN UV/optical continuum sampling ever obtained, with a mean sampling rate $<$0.5-day.
Approximately daily \hst\ UV sampling was also obtained. 
The UV/optical light curves show strong correlations ($r_\mathrm{max} = 0.57 - 0.90 $) and the clearest measurement to date of interband lags. 
These lags are well-fit by a $ \tau \propto \lambda^{4/3} $ wavelength dependence, with a normalization that indicates an unexpectedly large disk {\rb radius} of $\sim 0.35 \pm 0.05$~lt-day at 1367~\AA, assuming a simple face-on model.
The {\it U}-band shows a marginally larger lag than expected from the fit and surrounding bands, which could be due to Balmer continuum emission from the broad-line region as suggested by Korista and Goad.
The UV/X-ray correlation is weaker ($r_\mathrm{max} < 0.45$) and less consistent over time. 
This indicates that while \swift\ is beginning to measure UV/optical lags in {\rb general} agreement with accretion disk theory {\rb (although the derived size is larger than predicted), the relationship with} X-ray variability is less {\rb well} understood.
Combining this accretion disk size estimate with those from quasar microlensing studies suggests that AGN disk sizes scale approximately linearly with central black hole mass over a wide range of masses.
\end{abstract}


\section{Introduction}
\label{section:intro}

Because of their great distances and small sizes, the central regions of active galactic nuclei (AGN) cannot be resolved directly with current technology.
Thus it is necessary to use indirect methods to gain information about AGN structure and physical conditions.
Variability studies, along with gravitational microlensing (e.g., \citealt{Morgan10}, \citealt{Mosquera13}, \citealt{Blackburne14}, \citealt{Jiminez14}), have emerged as powerful techniques for probing the central {\rb regions} of AGN.

In particular, the ``reverberation mapping'' (RM) technique \citep{Blandford82} has proven quite effective at taking advantage of strong AGN line and continuum variability to probe the structure of the {\rb broad emission line region (BLR)}.
The fundamental idea of RM is that if the {\rb variability} in band B is powered by {\rb variability in} band A, with only light travel times affecting the light curves, then variations in band A will be seen in band B, but delayed and smoothed by the size and geometry of the latter emitting region.
The first unambiguous application of RM came in an {\it IUE} campaign on the Seyfert 1 galaxy \ngc, which found that variations in the driving UV continuum (band A in this picture) were highly correlated with those in emission lines such as C~{\sc iv} line (band B).
The line variations lagged the continuum by $\sim$10 days, indicating that the C~{\sc iv}-emitting region was of order 10 lt-days in size \citep{Clavel91}.
Optical emission lines showed similarly strong correlation but with larger lags.
For example H$\beta$ showed a lag of $\sim$20 days \citep{Peterson91}, indicating a stratified BLR in which higher-ionization lines are formed closer to the central engine.
The distance estimate, when combined with the line width, allows estimation of the mass of the central black hole. 
For \ngc\ the current best mass estimate is $ M_\mathrm{BH} \sim 3.2 \times 10^7$~M$_\odot$ (\citealt{Denney10}, \citealt{Pancoast14}).
This technique is now a standard tool for AGN astronomy, yielding BLR size, stratification information and black hole mass estimates and physical conditions for $\sim$50 AGN (see, e.g., \citealt{Bentz15} for a recent compilation).
{\rb For a more extensive general discussion of BLR RM, please see the first paper in this series (\citealt{DeRosa15}; Paper I hereafter).}

The structure and physics of the central engine that produces the continuum emission is currently less well understood than the reverberation-mapped BLR.
Observations and accretion disk theory both {\rb suggest} that the inner accretion disk/corona region emits short wavelength continuum (X-ray, ultraviolet [UV hereafter] and much of the optical), which then illuminates and ionizes the gas in the more distant BLR and beyond.
The prevailing picture is that the black hole is surrounded by a small, hot ($T \sim 10^9$~K) and relatively spherical corona and a larger, cooler ($T_\mathrm{max} \sim 5 \times 10^5$~K) and relatively flat accretion disk (e.g., \citealt{Haardt91}).
Gravitational lensing studies also indicate that {\rb this putative} corona is small enough to be considered point-like relative to the disk ($\sim 5 R_\mathrm{Sch}$; \citealt{Dai10}, \citealt{Morgan12}, \citealt{Mosquera13}, \citealt{Blackburne14}, \citealt{Blackburne15}).
The energy released by the accretion process heats both the optically thick disk -- producing the thermal UV/optical emission -- and the corona, which in turn can illuminate the disk as an external heating source. 
The fraction of the energy that goes into heating the corona has not been established and therefore it is not clear whether the disk is mainly heated internally or externally.
In either case, however, the disk is expected to have a stratified temperature structure with the hotter, UV-emitting regions closer in and the cooler, optically-emitting regions farther out.
Quasar microlensing studies find that accretion disk sizes increase with wavelength (e.g., \citealt{Poindexter08}), supporting this picture.
{\rb We note however that this picture contains important unreconciled discrepancies.
For instance gravitational lensing disk sizes are typically reported to be a factor of $\sim$4 larger than predicted (e.g., \citealt{Morgan10}), and the observed UV spectrum is too steep with a Lyman discontinuity that is typically smaller than predicted or not seen (e.g. \citealt{Koratkar99}, \citealt{Collin01}).
Recent improvements in AGN accretion disk models (e.g. \citealt{Dexter11}) may overcome these difficulties, but see also \cite{Antonucci13}.}

Just as RM of the BLR allows us to estimate the distances at which each line is formed, RM of the accretion disk could allow us to constrain the temperature structure of the disk and test the standard $\alpha$-disk model (\citealt{Shakura73} or any other predictive model).
Repeated efforts have been made to implement RM of the accretion disk by correlating X-ray light curves gathered with space-based observatories with optical light curves typically from ground-based observatories (e.g., \citealt{Edelson96}, \citealt{Nandra98}, \citealt{Suganuma06}, \citealt{Arevalo08}, \citealt{Arevalo09}, \citealt{Breedt09}, \citealt{Breedt10}, \citealt{Cameron12}, \citealt{Gliozzi13}).
However the optical time resolution of these early experiments was typically limited by the diurnal cycle to $ \Delta T > 1 $~day, resulting in lag measurements that were suggestive but not statistically significant ($ 1-2 \sigma $), although often in the expected sense, with X-rays leading the optical.
Other experiments from this period used \hst\ \citep{Edelson00} or {\it XMM-Newton} \citep{Mason02} to attain finer optical time resolution at the cost of shorter monitoring periods ($\sim$1-2~days).
Again the results were suggestive but inadequate to make a definitive lag measurement.
Further, these experiments typically only sampled a single optical or UV band, and thus were unable to explore temperature stratification in the disk.
Ground-based multicolor optical/IR studies also yielded tentative evidence of shorter wavelength variations leading longer wavelength variations in some AGN (\citealt{Sergeev05}, \citealt{Cackett07}, \citealt{Lira11}).

The unique capabilities of the \swift\ observatory \citep{Gehrels04}, originally optimized to detect $\gamma$-ray burst counterparts, are also ideally suited for AGN monitoring.
Its rapid slew/acquisition times and large sky coverage make it feasible to sample AGN light curves (which show variability over a broad range of temporal frequencies) with high cadence over a long duration.
Further, the coaligned UltraViolet-Optical Telescope (UVOT; \citealt{Roming05}) and X-Ray Telescope (XRT; \citealt{Burrows05}) cover the entire energy range of interest (the X-ray/UV/optical) with a single space-based telescope, so data quality is no longer limited by the diurnal cycle or weather.

This is leading to important advances in accretion disk RM, as highlighted by the success of two recent \swift\ AGN monitoring campaigns.
After detecting from the ground that the relatively normal galaxy NGC~2617 had transitioned into a Seyfert~1, \cite{Shappee14} used \swift\ to cover a $\sim$50~day period with approximately daily cadence, generally in all six UVOT filters.
Ground-based optical/infrared coverage of the first part of this period was obtained at a lower cadence.
\cite{McHardy14} analyzed  359 ``visits'' (separated by 1 orbit or longer; see Section 2.1) to the archetypical Seyfert 1 galaxy \ngc, over $\sim$2 years (2012 Feb - 2014 Feb).
Approximately 20\% of the visits utilized all six filters.
In both cases, the data show significant ($>3\sigma$) interband lags throughout the UV/optical, with increasing lags to longer wavelengths, consistent with a $\lambda^{4/3}$ dependence as predicted by the standard $\alpha$-disk model under the assumption that time lags are dominated by light travel times.

The experiment detailed herein combines \swift's {\rb powerful} capabilities with simultaneous, intensive UV spectroscopic monitoring by \hst\ to yield the densest X-ray/UV/optical coverage -- in both time and wavelength -- ever obtained.
\ngc\ is the target of this campaign.
Figure~1 shows that this campaign yields a factor of $\sim$2.5-4 improvement in the number of UVOT filter data pairs available for correlation compared to the two best previous campaigns.
This provides superior power to measure small ($<$2-day) interband lags with high precision.

The result, presented in this paper, is a clear measurement of lags across the entire UV/optical range, with shorter wavelength bands leading the longer wavelength bands.
The timescales generally increase to longer wavelengths as expected for a standard $\alpha$-disk, but direct fitting indicates a larger than expected disk ($\sim 0.35 \pm 0.05$~lt-day at 1367~\AA).
The {\it U}-band lag is slightly longer than expected from the fits, apparently consistent with contamination from BLR continuum emission as predicted by \cite{Korista01}.
The X-rays show a relatively weak and less coherent relation to the UV/optical.
Finally combining RM and microlensing disk size estimates suggest that disk size scales roughly linearly with black hole mass over a wide range of masses.

This paper is organized as follows.
Section 2 describes the observations and data reduction, 
Section 3 presents cross-correlation analyses applied to these data, 
Section 4 discusses the theoretical implications of these results, 
and Section 5 concludes with a brief summary of this work and implications for the future.

\section{Observations and data reduction}
\label{section:data}

\subsection{Observations}
\label{section:obs}

The target of this experiment, \ngc\ ($z = 0.01717$; \citealt{deVaucoulers91}), shows strong, reliable variability across the entire X-ray/UV/optical wavelength range accessible to \swift.
It is also among the brightest AGN in the sky at these wavelengths.
In 2014 February - June, \swift\ executed a monitoring campaign on \ngc\ that was ground-breaking in two respects: 
1) it was comprised of 360 separate visits over a $\sim$4 month period, of which 282 successful visits were obtained, for a sampling rate (after removing bad data) better than one visit every $\sim$0.5-day, and 
2) it utilized all six UVOT filters \citep{Poole08} in each visit, with 239 (84\%) providing usable measurements in all six filters.
(For the purposes of this paper, a visit is defined as an observation in which at least one UVOT filter measurement is obtained.
Multiple observations within a single $\sim$96~min orbit are combined to form a single visit.)
This entailed a significant commitment of spacecraft resources given the limit of 500 time-critical non-GRB guest investigator visits per year and the desire to minimize wear on the filter wheel.\footnote{http://swift.gsfc.nasa.gov/proposals/tech\_appd/swiftta\_v11\protect\linebreak[1]/node42.html}

In addition, a parallel \hst\ emission-line RM campaign yielded daily UV spectroscopic monitoring of \ngc\ over a slightly longer period (see Paper~I).
This provided mutual synergies: the \hst\ 1367~\AA\ continuum light curve was used in the cross-correlation functions (CCFs) reported herein, while the \swift\ optical, ultraviolet and X-ray light curves can be used to better define the continuum variability characteristics needed to understand the emission-line RM results.

These observations are summarized in Table~1.
Start and stop times for \swift\ observations are originally recorded in MET (Mission Elapsed Time; seconds since the start of 2001) and corrected for the drift of the on-board \swift\ clock and leap-seconds.
These times were averaged and converted to Heliocentric Julian Date (HJD), the standard for this observing campaign.
Throughout this paper we utilize the truncated HJD, defined as $ {\rm THJD = HJD - 2,456,000} $.
We reduced all \swift\ data on \ngc\ for both the UVOT and XRT, but restricted scientific analysis to observations taken during the intensive monitoring period, from THJD 706 to THJD 831
{\rb (approximately 2014 Feb 17.5 - Jun 22.5 UTC).}

\begin{deluxetable}{lcccc}
\label{table1}
\tablenum{1}
\tablecaption{Monitoring information}
\tablewidth{0pt}
\tablecolumns{5}
\tablehead{
\colhead{(1)} & \colhead{(2)} & \colhead{(3)} & \colhead{(4)} & \colhead{(5)}  \cr
 \colhead{} & \colhead{Central} & \colhead{Wavelength} & 
  \colhead{Number} & \colhead{Sampling}  \cr
 \colhead{Band} & \colhead{$\lambda$ (\AA)} & \colhead{range (\AA)} & 
   \colhead{of points} &  \colhead{Rate (day)} \cr}
\startdata						
 HX	  &	4.4	  &	1.2	-	15.5	  &	272	&	0.46	\cr
 SX  	&	25.3	&	15.5	-	41.3	&	272	&	0.46  \cr
 HST	&	1367	&	1364.5	-	1369.5	&	121	&	1.03	\cr
 UVW2	&	1928	&	1650	-	2250	&	262	&	0.47	\cr
 UVM2	&	2246	&	2000	-	2500	&	254	&	0.49	\cr
 UVW1	&	2600	&	2250	-	2950	&	266	&	0.47	\cr
 U	  &	3465	&	3050	-	3900	&	266	&	0.47	\cr
 B  	&	4392	&	3900	-	4900	&	265	&	0.47	\cr
 V  	&	5468	&	5050	-	5800	&	258	&	0.48	\cr
\enddata
\tablecomments{Column 1: Observing band name.
Column 2: Central wavelength of that band.
Column 3: FWHM wavelength range of that band, estimated from \cite{Poole08}.
Column 4: Total number of good data points in that band.
Column 5: Mean sampling rate in that band.}
\end{deluxetable}

The \hst\ 1367~\AA\ data reduction is detailed in Paper~I.
The following two subsections will describe the reduction of the \swift\ UVOT and XRT data.

\subsection{UVOT data reduction}
\label{section:uvot}

\swift\ observed \ngc\ for a total of 2935 exposures in six UVOT filters from the beginning of the mission through THJD 876.
All UVOT data were reprocessed for uniformity, applying standard {\tt FTOOLS} utilities (\citealt{Blackburne95}; from version 6.15.1 of {\tt HEASOFT}\footnote{http://heasarc.gsfc.nasa.gov/ftools/}).
The astrometry of each field was refined using up to 35 isolated field stars drawn from the \hst\ GSC 2.3.2 \citep{Lasker08} and Tycho-2 \citep{Hog00} catalogs, yielding residual offsets that were typically $\sim$0.3 arcsec.  
Fluxes were measured using a 5 arcsec circular aperture and concentric 40-90 arcsec regions were used to measure the sky background level. 
The final values include corrections for aperture losses, coincidence losses, and variation in the detector sensitivity across the image plane.
The galaxy contributes a fraction of the observed flux within the UVOT apertures (see Section~4.4) but no attempt was made to remove the contribution of host galaxy flux, as this contamination is constant and will not affect measurement of interband temporal correlations or absolute variability amplitudes.

We screened the data to eliminate exposures affected by significant tracking errors.  
To identify observations with distortions in the wings of their point spread function (PSF), we measured the ratio of counts in annuli from 5-7 and 7-10 arcsec, determined the distribution of these ratios for each filter and discarded any observations that were found to be outliers by at least 3.5$\sigma$ (defined iteratively).
In addition, we measured the PSFs of the isolated field stars used for astrometric refinement, flagging any observations for which either the average PSF Full width at half maximum (FWHM) differed by more than 1.0 arcsec from the nominal UVOT FWHM (2.2--2.9 arcsec depending upon the filter, \citealt{Breeveld10}) or the average FWHM of the stellar PSF projections along the X and Y axes differed by more than 0.75 arcsec.  
All flagged observations were manually inspected, leading to the rejection of one additional exposure in which the stars were streaks 15 arcsec long.
In total, 30 exposures are rejected.

The resulting light curves exhibited occasional, anomalously low points, especially in the UV.  
{\rb Subsequent investigation found that t}hese ``dropouts'' occur when the source falls within specific regions of the detector.  
Data potentially affected by these suspect detector regions are identified and removed using a new methodology discussed in the Appendix, eliminating 7.4\% of the exposures.
Finally we combined fluxes and errors in quadrature so there is no more than one data point per filter per orbit for any orbit in which multiple measurements were made in the same filter.
The final light curves are presented in Figure~2 and data from the full mission are given in Table~2.

\begin{deluxetable}{lccc}
\label{table2}
\tablenum{2}
\tablecaption{UVOT data}
\tablecolumns{4}
\tablehead{
\colhead{(1)} & \colhead{(2)} & \colhead{(3)} & \colhead{(4)} \cr
\colhead{HJD} & \colhead{Filter} & \colhead{Flux} & \colhead{Error} \cr}
\startdata
2454270.833	& UVW2 & 0.625 & 0.018	\cr
2454270.905	& UVW2 & 0.632 & 0.015	\cr
2454276.539	& UVW2 & 0.671 & 0.017	\cr
2454276.606	& UVW2 & 0.663 & 0.017	\cr
2454283.359	& UVW2 & 0.701 & 0.019	\cr
2454283.427	& UVW2 & 0.701 & 0.019	\cr
2454283.492	& UVW2 & 0.707 & 0.020	\cr
2454290.329	& UVW2 & 0.802 & 0.023	\cr
2454290.378	& UVW2 & 0.796 & 0.020	\cr
2454290.445	& UVW2 & 0.796 & 0.020	\cr
\enddata
\tablecomments{Column 1: Heliocentric Julian Date.
Column 2: Observing Filter.
Column 3: Measured flux in units of $10^{-14}$ erg cm$^{-2}$ s$^{-1}$ \AA$^{-1}$.
Column 4: Measured 1$\sigma$ error in the same units.
Note that this table includes all usable \swift\ observations of \ngc, not just those from the intensive monitoring period.
The data are sorted first by filter, then by HJD.
Only a portion of this table is shown here to demonstrate its form and content. 
A machine-readable version of the full table is available online.}
\end{deluxetable}

\subsection{XRT data reduction}
\label{section:xrt}

The \swift\ XRT data were gathered in photon counting (PC) mode and analyzed using the tools described by \cite{Evans09}\footnote{http://www.swift.ac.uk/user\_objects.} to produce light curves which are fully corrected for instrumental effects such as pile up, dead regions on the CCD and vignetting.
We generated soft X-ray (SX; 0.3--0.8 keV) and hard X-ray (HX; 0.8--10 keV) light curves. 
We utilized ``snapshot'' binning, which produces one bin for each spacecraft orbit. 
As with the UVOT data, we averaged multiple ObsIDs within a single orbit in quadrature. 
We investigated the use of other bands by subdividing HX into 0.8--2.8~keV and 2.8--10~keV but made no change after finding the correlation properties of the sub-bands to be very similar to the original choice.

The gap in the X-ray light curves during THJD 812-819 (2014 Jun 4--10; Figure~2) corresponds to the time that the Swift XRT was in an anomaly state (\citealt{Burrows14a}, \citealt{Burrows14b}, \citealt{Kennea14}), during which time XRT was either disabled or collected data in a non-standard, not-fully calibrated mode.
We excluded the data taken during this time interval from our analysis.
We additionally excluded all visits where the total good integration time was less than 120 sec.
This resulted in a final light curve having 272 X-ray points over the 125-day intensive monitoring period (see Table 1).
The complete \ngc\ XRT data are presented in Table~3.

\begin{deluxetable}{lcccc}
\label{table3}
\tablenum{3}
\tablecaption{XRT data}
\tablecolumns{5}
\tablehead{
\colhead{(1)} & \colhead{(2)} & \colhead{(3)} & \colhead{(4)} & \colhead{(5)} \cr
\colhead{HJD} & \colhead{HX Flux} & \colhead{HX Error} 
 & \colhead{SX Flux} & \colhead{SX Error} \cr}
\startdata
2453468.872	 &	0.324	 &	0.039	 &	0.093	 &	0.023	\cr
2453469.005	 &	0.361	 &	0.051	 &	0.061	 &	0.021	\cr
2453469.139	 &	0.282	 &	0.049	 &	0.156	 &	0.042	\cr
2453470.283	 &	0.425	 &	0.057	 &	0.114	 &	0.030	\cr
2453470.349	 &	0.343	 &	0.037	 &	0.075	 &	0.019	\cr
2453470.418	 &	0.360	 &	0.028	 &	0.076	 &	0.013	\cr
2453470.814	 &	0.353	 &	0.024	 &	0.088	 &	0.012	\cr
2453470.882	 &	0.469	 &	0.043	 &	0.088	 &	0.019	\cr
2453473.227	 &	0.496	 &	0.055	 &	0.175	 &	0.033	\cr
2453475.169	 &	0.384	 &	0.049	 &	0.073	 &	0.021	\cr
\enddata
\tablecomments{Column 1: Heliocentric Julian Date.
Columns 2 and 3: Measured HX flux and 1$\sigma$ error, in ct sec$^{-1}$.
Columns 4 and 5: Measured SX flux and 1$\sigma$ error, in ct sec$^{-1}$.
Note that this table includes all usable \swift\ observations of \ngc, not just those from the intensive monitoring period, sorted by HJD.
Only a portion of this table is shown here to demonstrate its form and content. 
A machine-readable version of the full table is available online.}
\end{deluxetable}

\subsection{Light curves}
\label{section:lcs}

Although the \swift\ data for \ngc\ span many years, Figure~2 and Table~1 cover only the $\sim$125-day intensive monitoring period THJD 706-831.
The light curves are presented in order of descending frequency with the highest frequency band at the top and the lowest at the bottom.
The \hst\ light curve plays a critical role  as the only data set not gathered by \swift.
This means that CCFs relative to this band will not suffer from ``correlated errors'' (see \citealt{Edelson88}).
The \hst\ light curve also has much higher signal-to-noise ratios and better exclusion of BLR emission than the \swift\ data, but with less than half the sampling cadence.

\section{Interband correlation and variability analyses}
\label{section:corr}

In this section we estimate the interband correlation and lag between continuum bands.
Before performing these correlation analyses we detrended the data by subtracting a 30-day boxcar running mean.
This was done to remove long-term trends that could potentially degrade our ability to measure the expected small lags.

In all correlation analyses we reference the correlation of one band (the \hst\ band) relative to all other bands (the eight \swift\ UVOT and XRT bands), restricting our analysis to just the data shown in Figure~2.
We used the interpolated cross-correlation function (ICCF) as implemented by \cite{Peterson04}, to measure and characterize temporal correlations within these data.
These results are shown in Table~4 and discussed in the following subsection.
The second subsection reports the result of fits to these data, and the third describes our characterization of the variable UVOT spectral energy distribution (SED).

\begin{deluxetable}{lcc}
\label{table4}
\tablenum{4}
\tablecaption{Interband correlation coefficients and lags}
\tablecolumns{3}
\tablehead{
\colhead{(1)} & \colhead{(2)} & \colhead{(3)} \cr
\colhead{Band} & \colhead{$r_\mathrm{max}$} & \colhead{Lag (days)} \cr}
\startdata
HX  	&	0.35 & $-$0.66 $\pm$ 0.46	\cr
SX	  &	0.44 & $+$0.08 $\pm$ 0.52	\cr
HST	  &	1.00 & $+$0.00 $\pm$ 0.25	\cr
UVW2  &	0.90 & $+$0.40 $\pm$ 0.17	\cr
UVM2	&	0.87 & $+$0.35 $\pm$ 0.16	\cr
UVW1	&	0.85 & $+$0.61 $\pm$ 0.20	\cr
U	    &	0.81 & $+$1.35 $\pm$ 0.24	\cr
B	    &	0.74 & $+$1.23 $\pm$ 0.29	\cr
V	    &	0.57 & $+$1.56 $\pm$ 0.50	\cr
\enddata
\tablecomments{Column 1: Band for which correlation was measured relative to \hst\ 1367~\AA. 
Column 2: Maximum correlation coefficient.  
Column 3: Measured centroid lag and associated 1$\sigma$ error in days.}
\end{deluxetable}

\subsection{Correlation analysis}
\label{section:iccf}

The traditional CCF (e.g., \citealt{Jenkins68}) requires evenly sampled data, but most astronomical data are not evenly sampled.
The ICCF performs a piecewise linear interpolation in the reference (\hst) band with a user-defined interpolation step of 0.1~day, and then measures the correlation relative to the non-interpolated data in the other band.
These data are then shifted and correlated to build up the correlation function.
In this case, the \hst\ data, with an initial cadence of $\sim$1.1~day, are resampled to a grid with 0.1~day spacing, and then the CCF of each \swift\ light curve is measured relative to the \hst\ light curve.

The results are shown in Figure~3a.
The third panel shows the auto-correlation function (ACF) of the \hst\ data; all others are CCFs measured relative to the \hst\ light curve, so a positive lag indicates that variations in that band lag behind the \hst\ light curve.
Two points are clear.
First there is a tendency for peak lags to increase with wavelength:
the hard X-ray band shows a negative lag relative to \hst, the soft X-rays show approximately zero lag, the lags are positive and small within the \swift\ UV (longer wavelengths lag \hst)
and the lags are positive and larger between \hst\ and the \swift\ optical.
Second the strength of the correlation is larger {\rb between} the \hst\ and UVOT data (peak correlation coefficients $ r_\mathrm{max} =0.57 - 0.90  $) than between the \hst\ and XRT data ($ r_\mathrm{max} < 0.45 $).

In order to quantify the uncertainties on the interband lag estimates, we utilized the flux randomization/random subset selection (FR/RSS) technique of \cite{Peterson98} as modified by \cite{Peterson04} to produce the cross-correlation centroid distribution (CCCD), as shown in Figure~3b. 
{\rb FR/RSS  is a model-independent Monte Carlo technique that attempts to deal with both flux uncertainties in individual measurements and uncertainties due to sampling of the time series. 
In ``random subset selection'' (RSS), for a light curve of $N$ data points, one randomly selects $N$ data point without regard to whether a data point has been previously selected or not.
Thus, approximately $1/e$ of the original points in the light curve are not selected in a given realization, and the remaining points are selected one or more times. 
For data points selected $n$ times in a given realization, the uncertainty associated with the data point is reduced by $n^{-1/2}$. 
``Flux randomization'' (FR) consists of altering the observed flux by random Gaussian deviates whose standard deviation is equal to the flux uncertainty on the data point. 
The CCCD is built by combining the results from 2000 realizations, with results that are summarized in Table~4.

A number of factors can contribute to the widths of the histograms and thus the error estimates on the interband lags.
These include the sampling cadence and finite duration of the campaign, measurement errors, and deviations from the stationarity assumption implicit in the FR/RSS method.
An example of the second contribution could be the appearance of somewhat different lags at different epochs, as may be occurring with the BLR (see Paper I).
At present it is not possible to determine the relative contribution of each effect.}

\subsection{Lag-wavelength fits}
\label{section:fits}

{\rb As discussed in the introduction, the standard model predicts a relationship between lag and wavelength because the disk is expected to be hotter at smaller, inner radii and cooler at larger, outer radii.}
Following the analysis of \cite{McHardy14} and \cite{Shappee14}, Figure~4 presents the CCF lag ($\tau$) results as a function of wavelength ($\lambda$).
We fit the wavelength dependence of the lags with the function $ \tau = A + B((\lambda/\lambda_0)^C-1)$.
The top three sets of panels show the effect of restricting the fitting function by first setting $ C = 4/3 $ and then setting $ A = 0 $.
This yields only a slight increase in $\chi^2_\nu$, which is acceptable in all cases.
For instance the third panel has reduced $\chi^2$ of $\chi^2_\nu = 0.98$, corresponding to a probability value $ p = 0.45 $.
Thus we conclude these data are fully consistent with a single-parameter fit, $ \tau = B((\lambda/\lambda_0)^{4/3}-1)$.
The fit parameter $B$ gives an estimate of the size of the disk at $\lambda_0 = 1367$~\AA, the \hst\ reference wavelength, assuming a face-on geometry.
This is the most important result of this paper, discussed in detail in Section~4.1.

The bottom three sets of panels explore the effect of excluding particular bands from the fit.
The third panel shows the effect of excluding the \hst\ ACF lag, which of course should be identically zero.
This has no effect on the number of degrees of freedom as the fit parameter $A$ is dropped as well (as discussed above).

The fourth panel shows the additional effect of dropping the two X-ray lags, HX and SX.
This has essentially no effect on fit quality, which is not surprising because the correlation coefficients are low and the lags have by far the largest errors of any waveband.
That is, the X-ray variations do not show a clear, consistent relation to the UV/optical variations.
The implications of this divergence is explored in detail in Section~4.4.

Another new result of this experiment is that the {\it U}-band lag is consistently larger than predicted by the fits.
The fifth set of panels show that additionally excluding the {\it U}-band lag from the fit shown in the fourth panel greatly reduces the $\chi^2_\nu$ although, as mentioned earlier, the fits are acceptable in all cases.
This is discussed in Section~4.2.

\subsection{Spectral variability}
\label{section:spvar}

In this section we utilize the fact that emission from the AGN (central engine and surrounding regions) is variable while starlight from the underlying galaxy is not to separate these components and characterize the spectral shape of the AGN component.
For most analyses, emission from the underlying galaxy is a complication to be removed from the UV/optical SED before proceeding.
One way to do this is image decomposition, as has been performed for \ngc\ by \cite{Bentz09}, \cite{Bentz13} and \cite{Mehdipour15}.
Here we use an alternate approach to estimate the shape of the variable SED (although not its normalization) directly from \swift\ data alone.

We first filter the intensive monitoring data to include only orbits with observations in all six UVOT filters in order to obtain a uniform data set.
For each band, we {\rb next} measure the standard deviation of the flux ($\sigma$) and the mean error ($\bar{\epsilon}$) and then calculate the error-corrected standard deviation,
$ \sigma_C = \sqrt{\sigma^2 - \bar{\epsilon}^2} $, although the error correction was always  small, typically $\sim$1\%.
This provides a direct estimate of the intrinsic variability in each band but does not include the mean flux of the AGN.
We then take the logarithm of both quantities and fit a function of the form
$ \rm log_{10}(\sigma_C(\lambda)) = \alpha \log_{10}(\lambda) + \beta $,
which is equivalent to 
$ \sigma_C(\lambda) \propto \lambda^\alpha $, in order to measure the power-law slope $\alpha$ of the variable component. 
As shown in Figure~5, this fit yields $ \alpha = -1.88 \pm 0.20 $.
In order to estimate the intrinsic shape of the variable AGN SED, we perform the same exercise after first dereddening the data (assuming $E(B-V) = 0.017$; see Paper~I).
This yields a slope of $ \alpha = -1.98 \pm 0.20 $.
We note that this is consistent with predicted thin accretion disk slopes of $ \alpha = -2 $ to $-2.33$ \citep{Davis07}.

\section{Discussion}
\label{section:disco}

\subsection{Reverberation mapping of the accretion disk}
\label{section:disk}

The standard model of a geometrically thin, optically thick AGN accretion disk predicts that the disk will be hotter in the inner radii and cooler in the outer radii, with dependencies on the black hole mass (and thus the Schwarzschild radius) and Eddington ratio.
This is for instance quantified in Equation~ 3.20 of \cite{Peterson97},
\begin{equation}\label{eq:}
T(r) \approx 6.3 \times 10^5 \left({\dot{M} \over \dot{M}_\mathrm{Edd}}\right)^{1/4} 
 M_8^{-1/4} \left( { r \over R_\mathrm{S} } \right)^{-3/4} {\rm K} 
\end{equation}
where $T(r)$ is the temperature at radius $r$, 
$\dot{M} / \dot{M}_\mathrm{Edd}$ is the mass accretion rate divided by the Eddington rate, assuming a radiative efficiency of $ \eta = 0.1$,
$M_8$ is the black hole mass in units of $10^8 \mathrm{M_\odot}$, and
$R_\mathrm{Sch}$ is the Schwarzschild radius.

Combining Equation 1 with Wien's law ($ \lambda_\mathrm{max} = 2.9 \times 10^7 /T $, where $\lambda$ is measured in {\AA}ngstroms and $T$ in Kelvin) and the idea that the lags are dominated by light travel times from the center (so $ \tau = r / c $) yields the relation $ \tau \propto \lambda^{4/3} $.
{\rb Note that while this derivation was for a disk heated ``internally'' by viscous processes, the same $\lambda^{4/3}$ dependence will arise in a disk heated ``externally'' by the putative central corona.
This is shown in Equations 1 and 2 of \cite{Cackett07} and Equation 4.56 of \cite{Netzer13}.
However these derivations all assume a relatively flat disk.
If the disk is strongly warped or if the corona is distributed across the disk \citep{Dexter11}, then an externally-heated disk will be hotter at large radii, leading to a flatter lag-wavelength relation.
The consistency of} the observed $ \tau - \lambda $ relation with the predicted relation {\rb broadly supports the standard accretion disk temperature profile.}

Equation 1 can also be used to estimate source parameters under the simple assumption of a face-on disk in which each annulus at temperature $T(r)$ radiates all its luminosity at $ \lambda_\mathrm{max} $ as given above.
The observed relation $ \tau = B((\lambda/\lambda_0)^{4/3}-1)$ has a single free parameter $B \approx 0.35 $~day, which indicates, for a face-on disk, that an annulus of radius $\sim$0.35~lt-day radiates at $ T = 2.9 \times 10^7 / \lambda_0 = 2.2 \times 10^4 $~K for $ \lambda_0 = 1367 $~\AA.
This distance of 0.35~lt-day corresponds to $ r/R_\mathrm{Sch} = 90 $ for a $ 3.2 \times 10^7 M_\odot $ black hole.

A more realistic picture would account for the fact that each annulus radiates as a blackbody of temperature $T(r)$ instead of radiating at a single wavelength $ \lambda_\mathrm{max} $.
Accounting for that will yield a larger value for the radius at which the disk emission peaks at $\lambda_\mathrm{max}$ because more flux at a given wavelength is produced by hotter blackbody emission interior to radius $R$ than by cooler blackbody radiation exterior to $R$.
A future paper in this series (Starkey et al., in prep.) utilizes direct modeling of each UV/optical light curve to produce a more rigorous analysis.

Nonetheless these large sizes may cause problems for the standard \cite{Shakura73} $\alpha$-disk model.
Assuming a value for $ \dot{M} / \dot{M}_\mathrm{Edd} = L/L_\mathrm{Edd} = 0.03 $, derived using a disk luminosity $L$ that is $\sim$37\% of the total luminosity, yields an accretion disk radius of only $ r/R_\mathrm{Sch} = 40 $ using the same formula and assumptions as above.
We note that there are models that produce effectively larger disks which can potentially better explain the UV/optical variability properties of AGN, such as the inhomogeneous disk model of \cite{Dexter11}.

\subsection{Contribution of BLR emission}
\label{section:BLR}

Another interesting result shown in Figure~4 is the longer {\it U}-band lag, relative to the fit and to the lags of nearby bands.
Excluding the {\it U}-band data from the fit yielded a significant improvement in $\chi^2_\nu$, although the overall fit is acceptable in either case.
The final fit on the bottom of Figure~4, which excludes {\it U}-band, predicts a {\it U}-band lag of $ \tau = 0.85 $ while the observed value is $ \tau = 1.35 \pm 0.24 $, a difference of 2$\sigma$.
In retrospect, one can see in both the previous \ngc\ campaign \citep{McHardy14} and the NGC~2617 campaign \citep{Shappee14} that the {\it U}-band lags were larger than the {\it B}-band lags, although those campaigns measured lags with much larger errors, so the deviation was not significant.
The vastly superior short timescale sampling provided by the current campaign (see Figure~1) allows for the measurement of this apparent effect with higher significance.

There is a simple explanation for this excess lag, discussed by \cite{Korista01}: Balmer continuum emission (both thermal diffuse and reflected incident continuum) and other pseudo-continuum emission from BLR clouds (e.g., UV Fe~{\sc ii}) contributes significantly to the observed {\it U}-band flux, and since the BLR is much larger than the optically-bright accretion disk, it will increase the observed lag.
This effect was seen by \cite{Maoz93} in \ngc\ RM data from the 1989 campaign.
The sensitivity of the strength of the diffuse continuum component to the presence of high gas densities and high ionizing photon fluxes make it an important diagnostic of the physical conditions within the BLR.
{\rb A possible alternative is that the Balmer continuum (and other pseudo-continua) is produced in an ``intermediate'' region smaller in size than the classical BLR but larger than the accretion disk.
This could more naturally explain the relatively small increment in the {\it U}-band lag, although it would also mean adding a previously-unknown emission component to the many already required to explain AGN spectral energy distributions.}

We note that continuum light curves measured at longer wavelengths and/or with narrower bands will be much less sensitive to this effect.
This cannot be done with \swift, but future papers in this series will analyze an expanded set of \hst\ and optical photometric bands (Fausnaugh \et in prep.) and ground-based spectroscopy of \ngc\ (Pei \et in prep.), providing a more sensitive test of the degree to which these CCFs are contaminated by {\rb emission from hot gas surrounding the central engine}.

\subsection{The accretion disk size - black hole mass relation}
\label{section:lens}

As discussed in Section~1, quasar gravitational microlensing studies have been used to estimate accretion disk sizes, finding a tendency for disk sizes to increase with black hole mass (\citealt{Morgan10}, \citealt{Mosquera13}).  
Disk RM measurements of Seyfert galaxies can be used to extend such relations to lower masses, luminosities and (probably) Eddington ratios, generally with smaller uncertainties because of the greatly reduced physical complexity of the measurement.  Figure~6 shows a summary of microlensing sizes estimates (the half-light radius $R_{1/2}$ at rest frame 2500\AA) from \cite{Mosquera13} as open triangles.  
Microlensing studies frequently focus on $R_{1/2}$ because estimates of its value are relatively insensitive to changes in the underlying (disk) emission
profile (\citealt{Mortonson05}).  
A fit to these data as a power-law, $ \log_{10}(R_{1/2}) = A + B \log_{10}(M/M_0) $,
with $M_0 \equiv 3 \times 10^8 M_\odot$ to minimize covariances between the parameter estimates and assuming $0.3$~dex uncertainties in the black hole mass estimates, yields $ A = \log (R_0/\hbox{cm}) = 15.81 \pm 0.16 $ and $ B = 1.29 \pm 0.33 $ with $ \chi^2 = 8.73 $ for 9 dof, a statistically acceptable fit.
The slope of the fit is driven by the higher mass systems, leading it to lie below the measurements in the mass range of \ngc.

The comparable disk RM size from the present study is the distance corresponding to the lag for the UVW1 filter centered at 2600\AA, which we show as the filled square labeled ``N5548''. 
This combines the parameter $B$ ($=0.35\pm0.04$~days for the simple face-on model in Figure~4) with the HST UV to UVW1 lag ($0.61\pm0.20$~days), to give an estimate that $R_{1/2}=0.96\pm0.21$~light-days.  
This assumes that the size corresponding to the observed lag corresponds to $R_{1/2}$, which may not be correct, but is a reasonable assumption pending a theoretical model for how the disk RM lag should be interpreted in detail.  
For example, $R_{1/2}$ is 2.44 times larger than the radius at which the photon wavelength matches the disk temperature discussed in Section~4.1.
A similar procedure was used to add the UVW1 size estimate for NGC~2617 from \cite{Shappee14}, where the error bar is designed to span their systematic uncertainties.

While there are residual systematic uncertainties in this comparison, such as the meaning of the disk RM lag as a physical size and potential differences in the Eddington ratios of the nearby lower luminosity Seyfert~1s and the distant high luminosity quasars, the results from the two very different methods are broadly consistent.  
If we simply fit the combined data, we find $ A = 15.96 \pm 0.12 $ and $ B = 0.98 \pm 0.23 $ with $ \chi^2 = 13.51 $ for 11 dof, which is shown by the dashed line in Figure~6.  
The slope is flatter and better defined, the fit is consistent with the combined data, and the parameters are consistent with the results using only the microlensing results.
Interestingly, the slope $B$ is almost exactly unity (and thus the fit line is nearly parallel with the last stable orbit, shown as a solid line in Figure~6), indicating that the disk size in units of Schwarzschild radii is nearly constant, $ R_{1/2}(2500\AA) \sim 100 R_{Sch} $, over a very wide range of AGN masses.
That these two radically different methods agree this well seems remarkable given the different underlying physics and the possible range of systematic effects.

\subsection{Relation of X-ray to UV/optical continua}
\label{section:xray}

Figure~3 shows that compared to the strong ($ r_\mathrm{max} = 0.57 - 0.9 $) correlations within the UV/optical, the correlation between the \hst\ 1367~\AA\ and the X-ray bands is much weaker ($ r_\mathrm{max} = 0.35-0.44 $).
This is surprising because it is well established that the optical and X-ray light curves of \ngc\ are very well correlated ($ r_\mathrm{max} = 0.95 $) on longer timescales of years \citep{Uttley03}.
That is, the strong long timescale optical/X-ray correlation does not translate to strong short timescale UV/X-ray correlations in \ngc.
Visual examination of the \ngc\ UV and X-ray light curves both in this paper and \cite{McHardy14} shows that there are some periods in which the UV appears to lead the X-rays, some in which the UV appears to lag the X-rays, and some in which there is no simple discernible relationship.

Periods of uncorrelated X-ray/optical variations are also seen in other Seyferts (e.g. NGC~3516, \citealt{Maoz02} and Mkn~79, \citealt{Breedt09}).
The phenomenon may be linked to internal heating fluctuations in the disk which are not `seen' by the X-ray emitting region, perhaps linked to mass accretion fluctuations which do not propagate to the central X-ray emiting region due to viscous damping (e.g., the explanation of a similar phenomenon seen on equivalent, mass-scaled time-scales in a stellar mass black hole X-ray binary \citealt{Cassatella12}).
This may indicate that a significant fraction of the UV/optical emission is not due to reprocessing of X-ray photons, but rather is generated internally.
In this case, the observed time lags would not be dominated by light travel effects, but instead would depend on the physics of the internal disk variations.

Alternatively, the lack of correlation between the UV/optical and X-ray could be a signature of absorption due to intervening material, but then one would expect the UV to be better correlated with the hard X-rays than the soft X-rays, because the latter would be much more strongly affected (and the light curves more decorrelated) by ''warm" absorption.
We do know that \ngc\ shows strong variable absorption in the X-rays \citep{Mehdipour15}, and there are indications that the absorption was changing (decreasing) during the \swift\ campaign.
Nonetheless the fact that the hard X-rays show a smaller correlation coefficient than the soft X-rays suggests that this may not be a complete explanation. 
A somewhat different scenario, based on a correlation between soft X-ray excess and
far-UV also observed by \citep{Mehdipour15}, is that both are associated with Comptonization.
Finally, it could simply mean that the observed 0.3-10 keV X-ray band is a poor proxy for emission from the {\rb putative} hot corona, which should emit the bulk of its luminosity at harder energies.
At this point it is not possible to say with certainty which, if any, of these explanations is correct.

\section{Conclusions}
\label{section:conclusions}

This paper presents the results of the most intensive X-ray/UV/optical AGN monitoring ever, spanning a duration of months.
We find that the UV/optical light curves are all well correlated with lags of $\sim$1-2 days increasing to longer wavelengths.
These lags are well-fitted by the relation $ \tau \propto \lambda^{4/3} $, in agreement with standard steady-state accretion disk predictions under the assumption that time lags are dominated by light travel times.
The fits yield a disk size of $\sim 0.35 \pm 0.05$~lt-day at 1367~\AA, larger than expected from standard $\alpha$-disk models or extrapolation from higher-mass microlensing studies.
Interestingly the {\it U}-band lag is anomalously large, suggesting that the {\it U} band is affected by Balmer continuum emission from the BLR.
The X-ray/UV correlations are weaker and less consistent, however, so these data do not confirm all predictions of the reprocessing picture.

We are planning a series of future papers to explore these results in greater detail.
Fausnaugh \et (in prep.) will present ground-based optical and further \hst\ continuum data, allowing a check on the wavelength dependence of the observed interband lags.
Pei \et (in prep.) will use ground-based spectroscopy to measure the continuum in narrow spectral windows much less affected by BLR emission, further refining this analysis.
Starkey \et (in prep.) will apply Markov Chain Monte Carlo methods to directly model disk emission from these continuum data, allowing a much more direct probe of the physical conditions.

This RM disk size estimate of a relatively low-mass Seyfert~1 galaxy forms a nice complement to more numerous but more uncertain accretion disk size estimates derived from generally higher-mass quasar microlensing studies.
The combination of the two datasets allows improved determination of the accretion disk size - black hole mass relation, which interestingly suggests 2500~\AA\ accretion disk sizes of $ R_{1/2} \approx 100 R_\mathrm{Sch}$.
Further Seyfert~1 accretion disk RM experiments will allow this relation to be tested and refined.

Most important for the long term is that this experiment demonstrates how dense and broad coverage in both wavelength and time can be used to probe a nearby AGN accretion disk with unprecedented detail.
\swift\ was named ``the premier facility for multi-wavelength time domain astronomy'' by the latest NASA Senior Review Panel.\footnote{http://science.nasa.gov/media/medialibrary/2014/05/15/\protect\linebreak[1]Final\_Report\_Astro2014\_SeniorReview\_Panel.pdf}
There is certainly no other observatory that can single-handedly monitor AGN with such dense and broad temporal and frequency coverage in the UV/optical and X-rays.
We expect that this experiment will become a template for future \swift\ campaigns that characterize the accretion disks of a sample of AGN covering a range of black hole masses, Eddington ratios and other source parameters.

\acknowledgments 
Support for HST program number GO-13330 was provided by NASA through a grant from the Space Telescope Science Institute, which is operated by the Association of Universities for Research in Astronomy, Inc., under NASA contract NAS5-26555.
{\rb RE gratefully acknowledges support from NASA under awards NNX13AC26G, NNX13AC63G, and NNX13AE99G.}
JMG gratefully acknowledges support from NASA under award NNH13CH61C.
BMP, GDR, CJG, MMF, and RWP are grateful for the support of the National Science Foundation through grant AST-1008882 to The Ohio State University. 
AJB and LP have been supported by NSF grant AST-1412693. 
MCB gratefully acknowledges support through NSF CAREER grant AST-1253702 to Georgia State University.
KDD is supported by an NSF Fellowship awarded under grant AST-1302093.
PBH is supported by NSERC. 
SRON is financially supported by NWO, the Netherlands Organization for Scientific Research.
BCK is partially supported by the UC Center for Galaxy Evolution.
CSK acknowledges the support of NSF grant AST-1009756.
PL acknowledges support from Fondecyt grant \#1120328.
AP acknowledges support from a NSF graduate fellowship and a UCSB Dean's Fellowship. 
JSS acknowledges CNPq, National Council for Scientific and Technological Development (Brazil) for the partial support and The Ohio State University for warm hospitality.
TT has been supported by NSF grant AST-1412315.
TT and BCK acknowledge support from the Packard Foundation in the form of a Packard Research
Fellowship to TT. 
TT thanks the American Academy in Rome and the Observatory of Monteporzio Catone for kind hospitality.
The Dark Cosmology Centre is funded by the Danish National Research Foundation.
MV gratefully acknowledges support from the Danish Council for Independent Research via grant no. DFF – 4002-00275.
This research has made use of the NASA/IPAC Extragalactic Database (NED), which is operated by the Jet Propulsion Laboratory, California Institute of Technology, under contract with the National Aeronautics and Space Administration.

\setcounter{table}{0}
\renewcommand{\thetable}{A\arabic{table}}


\appendix
\label{section:appendix}

As discussed in Section 2.2 we discovered ``dropouts'' in the UVOT light curve in the course of {\rb the} data reduction: isolated points with fluxes many sigma below those of their nearest neighbors.
Figure A1 shows the  UVOT light curves of \ngc\ after initial flux measurements and removal of data points affected by tracking problems.
The dropouts are the points with red error bars, most frequently seen in the UV bands.
In order to quantify this effect, we first parametrized the deviation for every point in the light curve as $ df = (F_N - 0.5 \times (F_{N-1}+F_{N+1}))/\sigma_N $, where $F_N$ and $\sigma_N$ are the measured flux and 1$\sigma$ error bar for visit $N$. 
To minimize the effects of intrinsic variability, we only tested data with observing gaps $ t_{N+1} - t_{N-1} < 2.5 $ days ($t_N$ is the THJD of the $N$th visit); all visits with longer gaps between their nearest neighbors were ignored. 

We then flagged all points with negative excursions greater than the largest positive excursion seen in that filter as dropouts.
The largest positive excursions used to define the threshold of what is a dropout are themselves sensitive to dropout measurements, in that the largest $df$ values tend to be found when the $N-1$ or $N+1$ flux measurement is a dropout.  
We therefore redefine these thresholds iteratively, removing the dropout points and then re-evaluating the largest positive excursions.  
This is repeated until the largest positive excursion remained the same, that is, it was not associated with a dropout.
This procedure limits the number of false positives to of order one per light curve. 
Note the strong dependence on UVOT band, with 33, 13, 21, 11, 2 and 1 dropouts for the {\it UVW2, UVM2, UVW1, U, B} and {\it V} bands, respectively.

We then mapped the source location of every exposure for the three UV filters ({\it UVW2, UVM2, UVW1}) to the UVOT detector coordinates (Figure A2). 
Blue dots show the points used in the analysis, red Xs the dropouts {\rb as defined above}, and open black circles the points ignored because they lacked nearby (in time) neighbors. 
Figure A3 is a blow-up of the region with the dropouts. 
Note that the dropouts cluster together such that most can be enclosed by a small number of boxes.  
We define eight rectangles in the detector plane, one for each cluster of at least three dropout points.
The coordinates of these boxes are given in Table~A1.
We then went back to all six filters (including the optical UBV bands) and flagged every visit/filter that falls in any of these boxes. 

\begin{deluxetable}{cccc}
\label{tablea1}
\tablenum{1}
\tablecaption{UVOT detector bad boxes}
\tablewidth{150pt}
\tablecolumns{4}
\tablehead{\colhead{x1} & \colhead{x2} & \colhead{y1} & \colhead{y2} \cr}
\startdata
357 &  371 &  634 &  651 \cr
415 &  440 &  582 &  632 \cr
441 &  469 &  634 &  655 \cr
448 &  457 &  440 &  442 \cr
479 &  484 &  539 &  542 \cr
516 &  537 &  596 &  619 \cr
536 &  538 &  564 &  568 \cr
545 &  583 &  573 &  607 \cr
\enddata
\tablecomments{UVOT detector coordinates (x1,x2,y1,y2) for the eight ``bad'' boxes shown in Figure~A3.
These are image coordinates for a full-frame raw UVOT image with 2$\times$ binning (the default UVOT image mode).}
\end{deluxetable}

These additional suspect points are shown as red Xs in Figure A1. 
We ran Kolmogorov-Smirnov tests on each band to test if the deviation for points in the suspect regions (the red Xs in Figure~A1) derives from the same population as the unaffected data (black dots in Figure~A1).
We find that for five of the six UVOT filters (all except {\it V}) the two samples are not consistent at the $10^{-4}$ to $10^{-16}$ level, with the strongest differences at the highest frequencies.
We then eliminated all data points in any filter that fell in these eight boxes.
These points are shown as red Xs in Figure A1.  
The remaining points (the black dots in Figure~A1) then formed the final light curve shown in Figure~2 of the main section of this paper.



\begin{figure}
\figurenum{1}
\begin{center}
 \includegraphics[angle=270,width=2.9in]{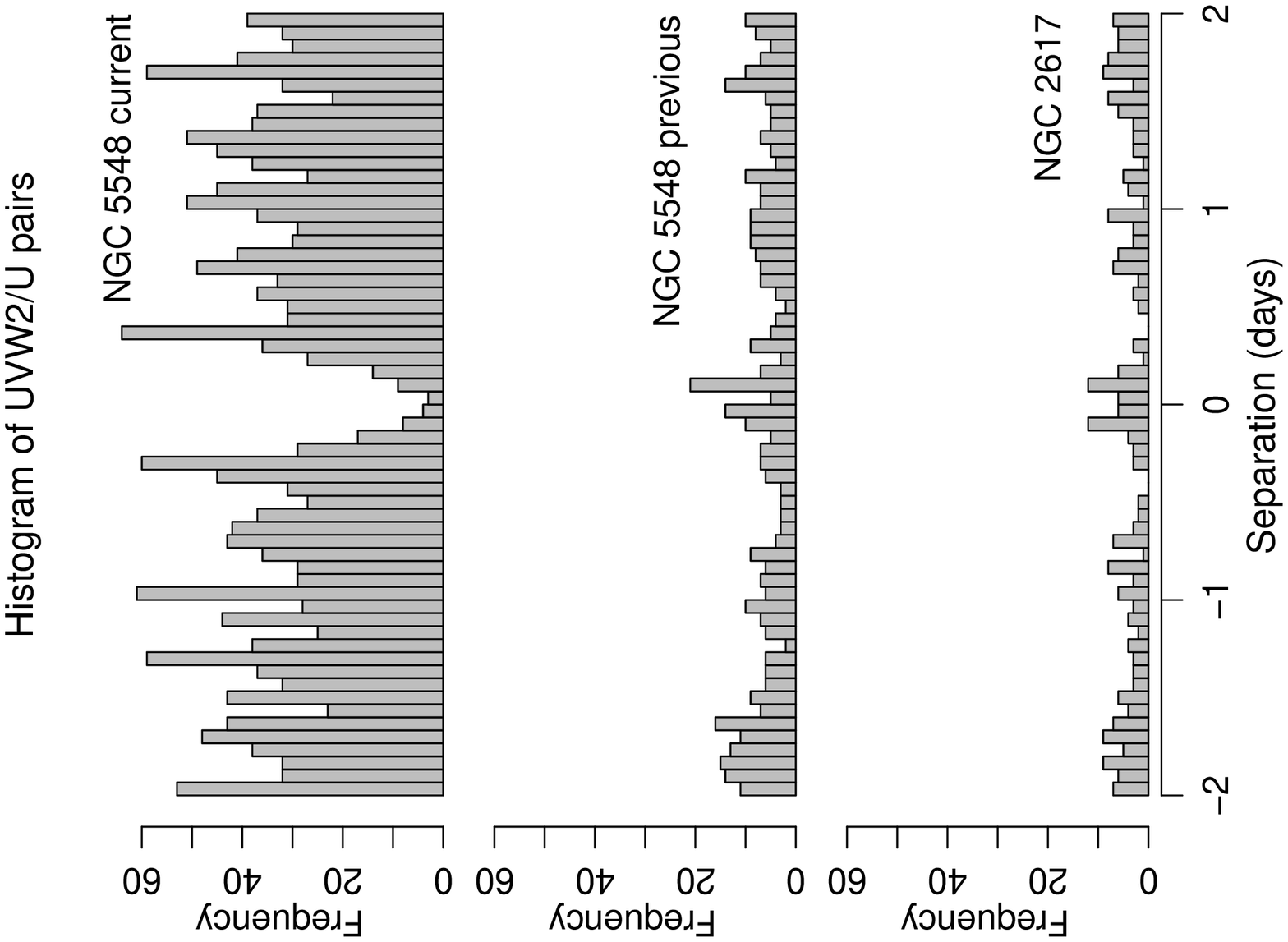} 
\caption{Histograms showing the number of {\it UVW2/U} pairs for  the current intensive \ngc\ monitoring campaign (top), the earlier \ngc\ campaign (middle, \citealt{McHardy14}) and the NGC~2617 campaign (bottom, \citealt{Shappee14}).
Data are binned by orbit, and all pairs with separations of less than half an orbit are excluded.
{\it UVW2} was used because it was the most frequently observed UVOT band, while {\it U}, a typical less well sampled band, was used because that band is particularly interesting (see Section~4.2).
The range $\pm$2~days is shown because this is the key cadence range that has not previously been well sampled.
Note that the current campaign samples these short cadences $\sim$5-8 times more frequently than the previous \ngc\ and NGC~2617 campaigns.}
\label{fig:fig1}
\end{center}
\end{figure}

\begin{figure}
\figurenum{2}
\begin{center}
 \includegraphics[angle=270,width=6.8in]{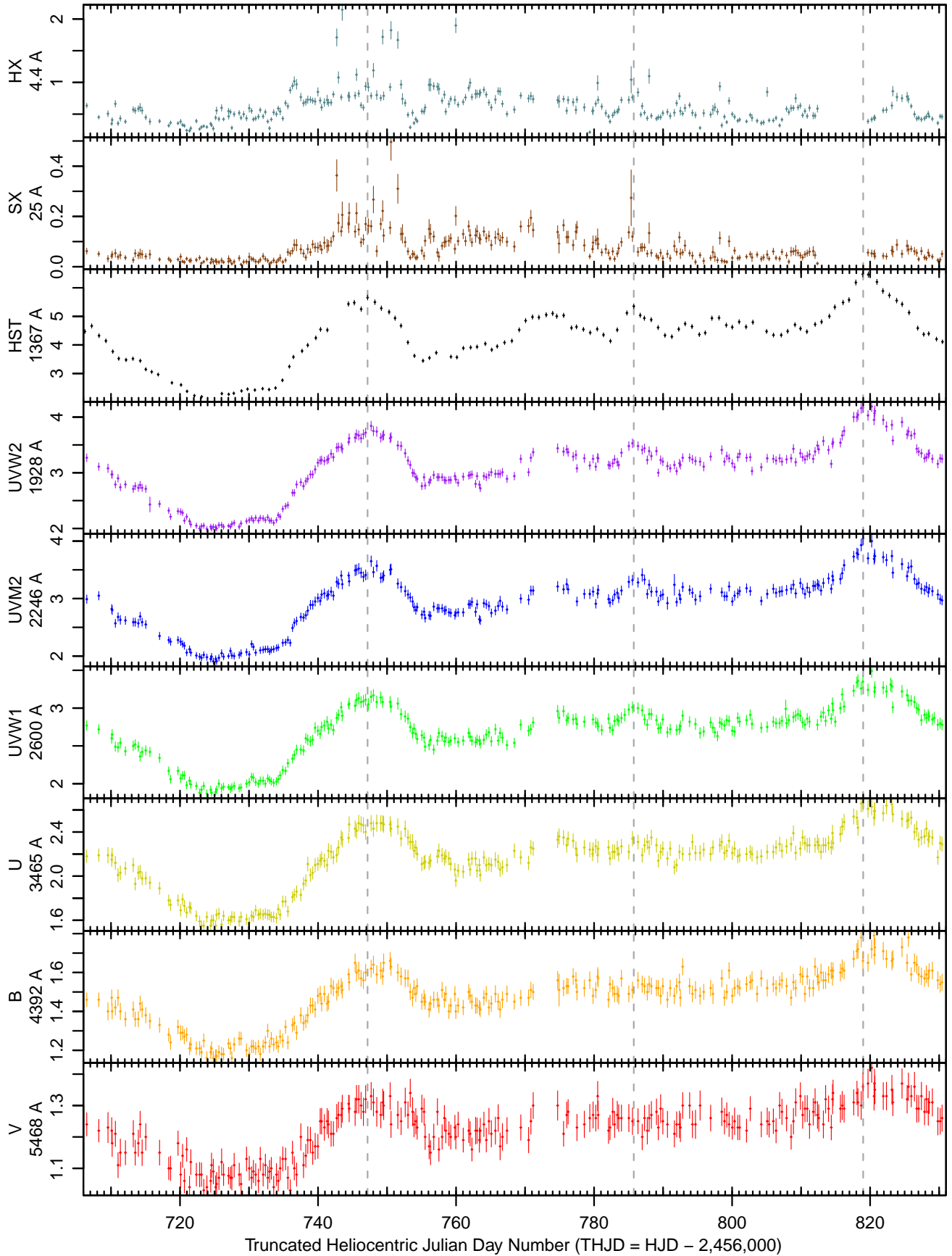} 
\caption{Light curves for the intensive monitoring period (HJD 2,456,706-2,456,831), going from shortest wavelength (top) to longest (bottom).
{\rb The band name and central wavelength are given on the left of each panel.}
Top two panels show the \swift\ hard and soft X-ray (HX and SX respectively) light curves, in units of c/s.
Third panel shows the \hst\ light curve, in units of $ \rm 10^{-14} erg s^{-1} cm^{-2} \AA^{-1} $.
Error bars for this light curve are typically $\sim$1.5\%, just barely visible in the plot.
The bottom six panels show the \swift\ light curves, again in units of $10^{-14}$ erg~cm$^{-2}$s$^{-1}$\AA$^{-1}$ .
Dashed gray lines show times THJD 747.179, 785.752 and 818.993, three local maxima of the \hst\ light curve.
}
\label{fig:fig2}
\end{center}
\end{figure}

\begin{figure}
\figurenum{3}
\begin{center}
\epsscale{1.0}
 \includegraphics[angle=-90,width=3.4in]{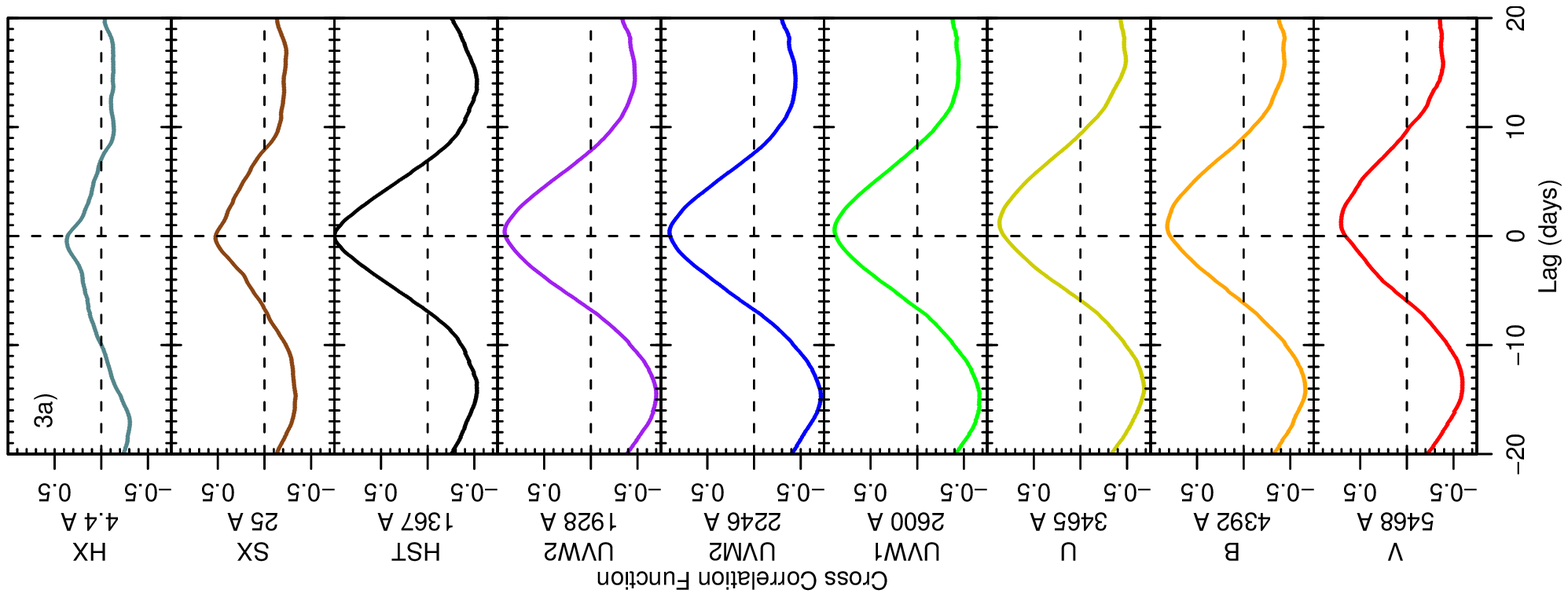} 
 \includegraphics[angle=-90,width=3.4in]{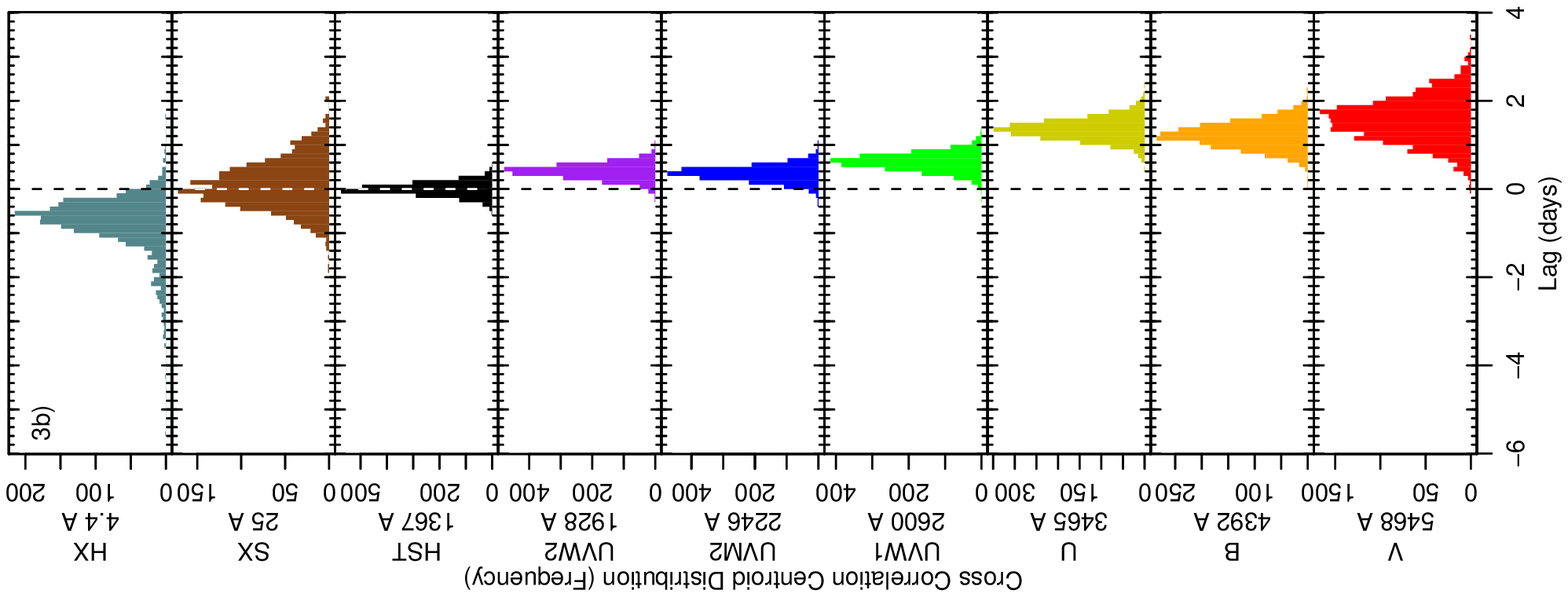} 
\caption{(3a) Interpolated cross-correlation functions for the intensive monitoring period light curves (Figure~2), with all correlations measured relative to the \hst\ light curve, after removing long term trending (see Section 3).
{\rb The band name and central wavelength are given on the left of each panel.}
Note that the interband lag goes from negative to increasingly positive as the band's wavelength increases.
Note also that the UV/optical correlations are all strong ($r_{max} = 0.57 - 0.90 $) but the X-ray/UV correlations are much weaker, ($r_{max} < 0.45$).
(3b) Cross-correlation centroid histograms derived from the CCFs as discussed in the text.
{\rb The band name and central wavelength are given on the left of each panel.}
All distributions except HX appear consistent with a Gaussian.}
\label{fig:fig3}
\end{center}
\end{figure}

\begin{figure}
\figurenum{4}
\begin{center}
\epsscale{1.0}
 \includegraphics[angle=-90,width=2in]{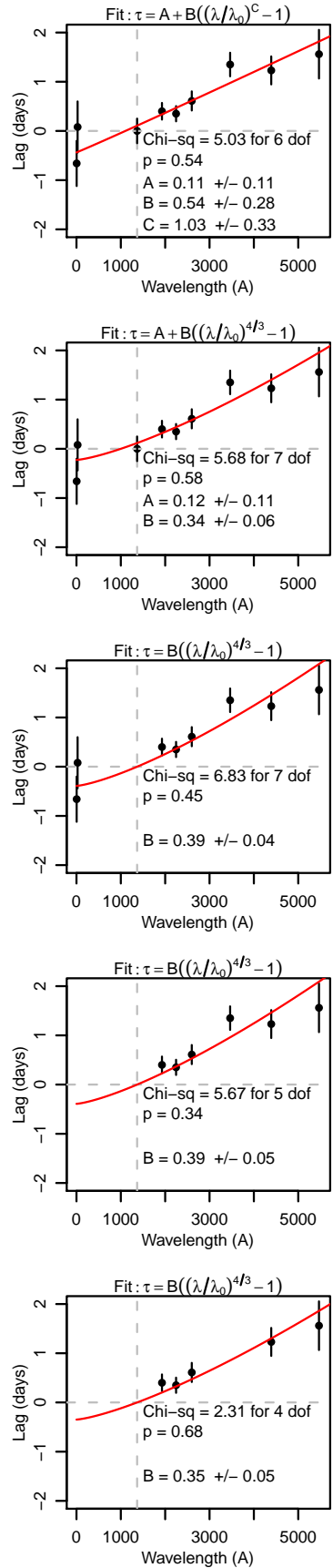} 
\caption{Lag-wavelength fits based on the data in Tables 1 and 4.
The top row shows the most general fit, $ \tau = A + B((\lambda/\lambda_0)^C-1)$, with the power-law index $C$ allowed to float.
The next row fixes the index at the theoretically expected value $ C = 4/3 $.
All data are included in the first two sets of fits.
In the third row the intercept is fixed at $ A = 0 $ and the \hst\ ACF data are excluded.
The fourth row shows these fits with the X-ray data HX and SX also excluded, while the fifth row additionally excludes the {\it U}-band data.
See text for further details.}
\label{fig:fig4}
\end{center}
\end{figure}

\begin{figure}
\figurenum{5}
\begin{center}
\epsscale{1.0}
 \includegraphics[angle=-90,width=3in]{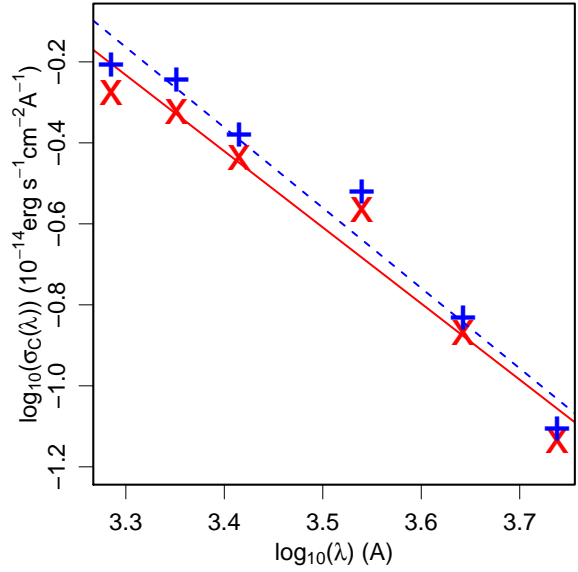} 
\caption{The error-subtracted variable flux ($\sigma_C(\lambda)$) as a function of wavelength ($\lambda$).
The original data are shown as red Xs and the dereddened data as blue crosses.
A fit to the function $ \rm log_{10}(\sigma_C(\lambda)) = \alpha \log_{10}(\lambda) + \beta $ yielded $ \alpha = -1.88 \pm 0.20 $ for the original data (solid red line) and $ \alpha = -1.98 \pm 0.20 $ for the dereddened data (dashed blue line).}
\label{fig:fig5}
\end{center}
\end{figure}

\begin{figure}
\figurenum{6}
\begin{center}
\epsscale{1.0}
 \includegraphics[angle=0,width=3in]{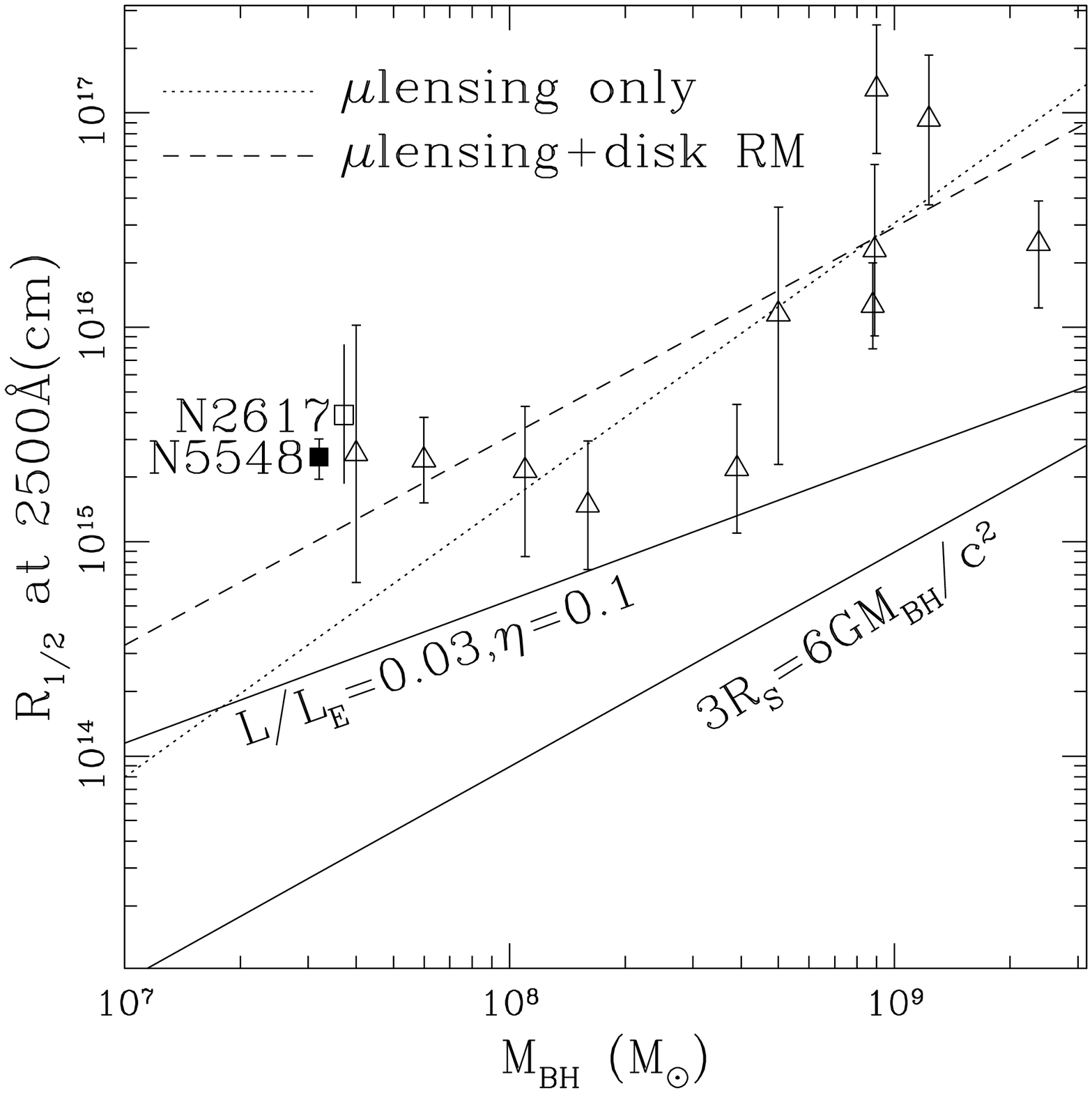} 
\caption{Accretion disk size estimates from quasar microlensing studies (open triangles, \citealt{Mosquera13}), the current study of \ngc\ (black filled square) and NGC~2617 (open square, \citealt{Shappee14}) as a function of black hole mass.
The dotted line shows the fit to just the microlensing data and the dashed line the fit to all data points including the Seyfert~1 RM measurements. 
To give a sense of other scales associated with accretion disks, the lower solid line shows the last stable orbit of a non-rotating black hole at $ R = 3R_\mathrm{Sch} = 6 GM/c^2 $ and the upper solid lines shows $R_{1/2}$ at 2500\AA\ for a simple thin disk with Eddington ratio $ L/L_\mathrm{Edd} = 0.03 $ and efficiency $\eta=0.1$ to provide a sense of scale.  
The latter curve can be shifted as $ (L/L_\mathrm{Edd})^{1/3} $ for different choices of these factors.}
\label{fig:fig6}
\end{center}
\end{figure}

\clearpage

\begin{figure}
\figurenum{A1}
\begin{center}
\epsscale{1.0}
 \includegraphics[angle=-90,width=7in]{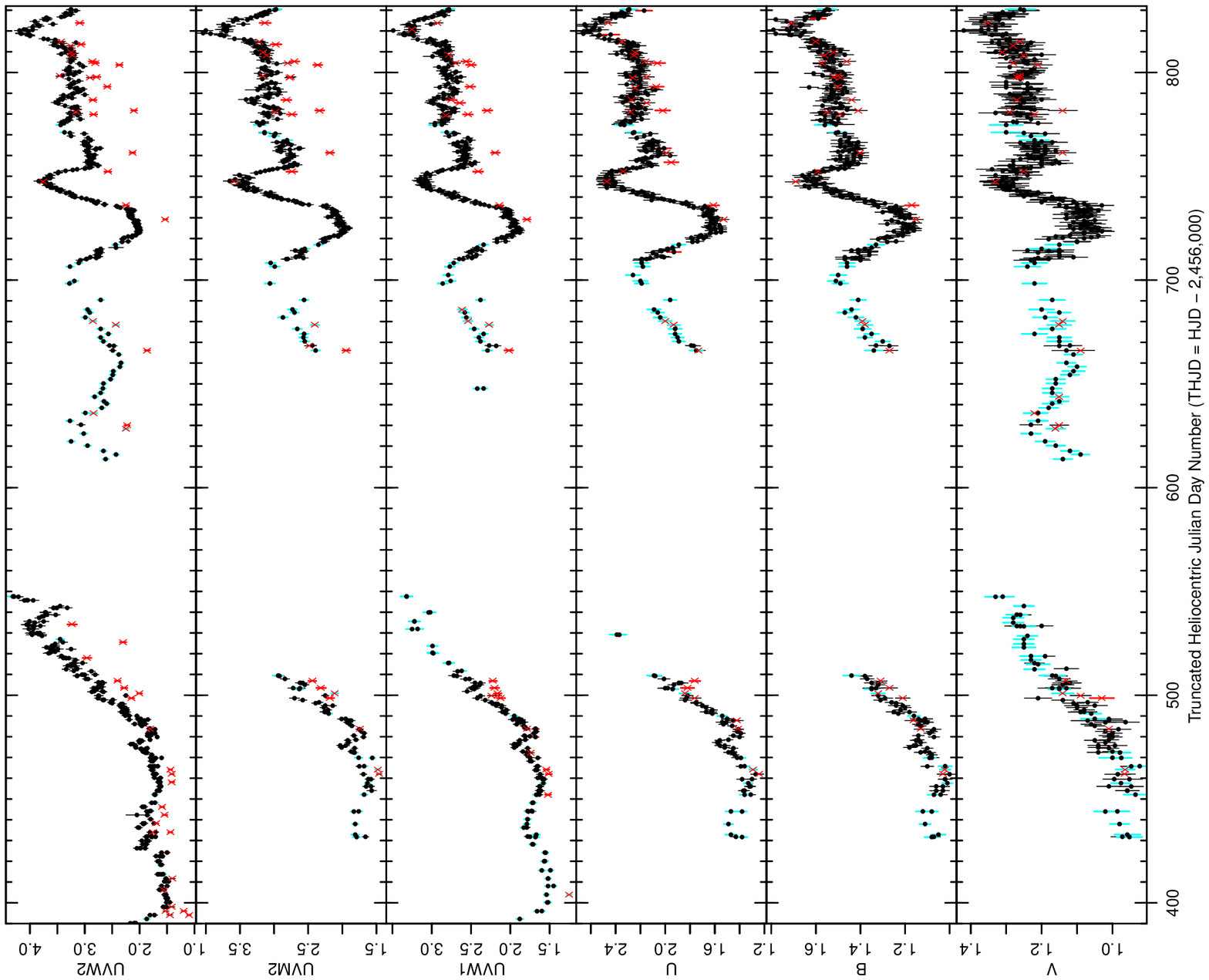} 
\caption{Initial UVOT light curves of \ngc\ for the period THJD 390-832.
The error bar colors indicate the results of the dropout test: 
black errors indicated that the point passed the dropout test (small deviation), red error bars indicated that it failed the test (large deviation) and cyan error bars indicate that it was not tested (as it lacked sufficient nearby neighbors). 
The symbols show if the point fell inside/outside the UVOT boxes (shown in Figure A3): black dots fell outside the boxes and thus were used in the final light curve (shown in Figure 1) and red Xs fell inside the boxes and were excluded from the final light curves.}
\label{fig:figa1}
\end{center}
\end{figure}

\begin{figure}
\figurenum{A2}
\begin{center}
\epsscale{1.0}
 \includegraphics[angle=-90,width=3in]{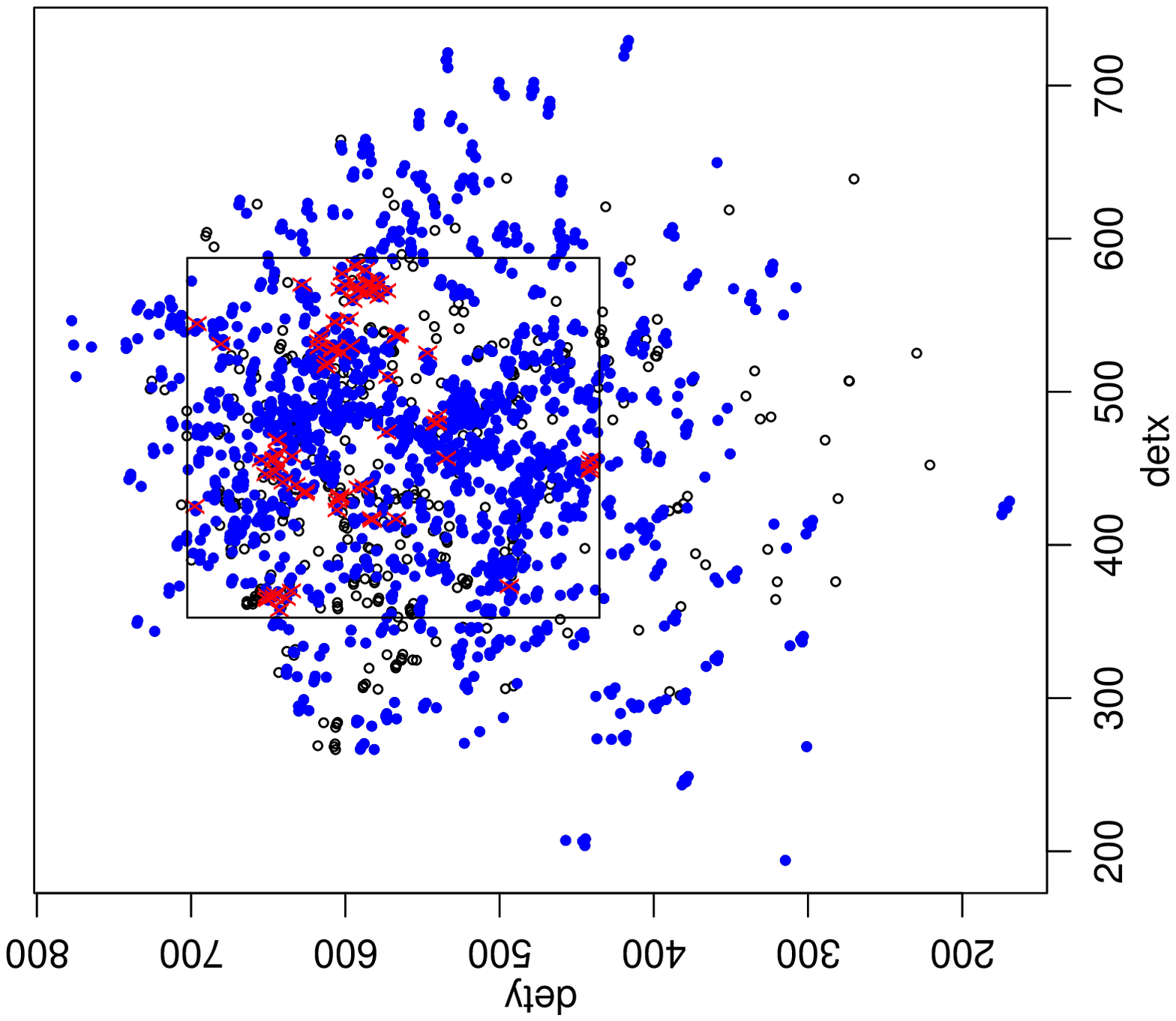} 
\caption{UVOT detector coordinates of the {\it UVW2, UVM2} and {\it UVW1} data in Figure~A1.
Data that were tested for dropouts are shown as blue dots and those that were not tested (due to their not having sufficiently nearby neighbors) are shown as open black circles.
The points that failed the dropout test are marked with red Xs.
The black rectangle denotes the region shown in Figure A3.}
\label{fig:figa2}
\end{center}
\end{figure}

\begin{figure}
\figurenum{A3}
\begin{center}
\epsscale{1.0}
 \includegraphics[angle=-90,width=4.1in]{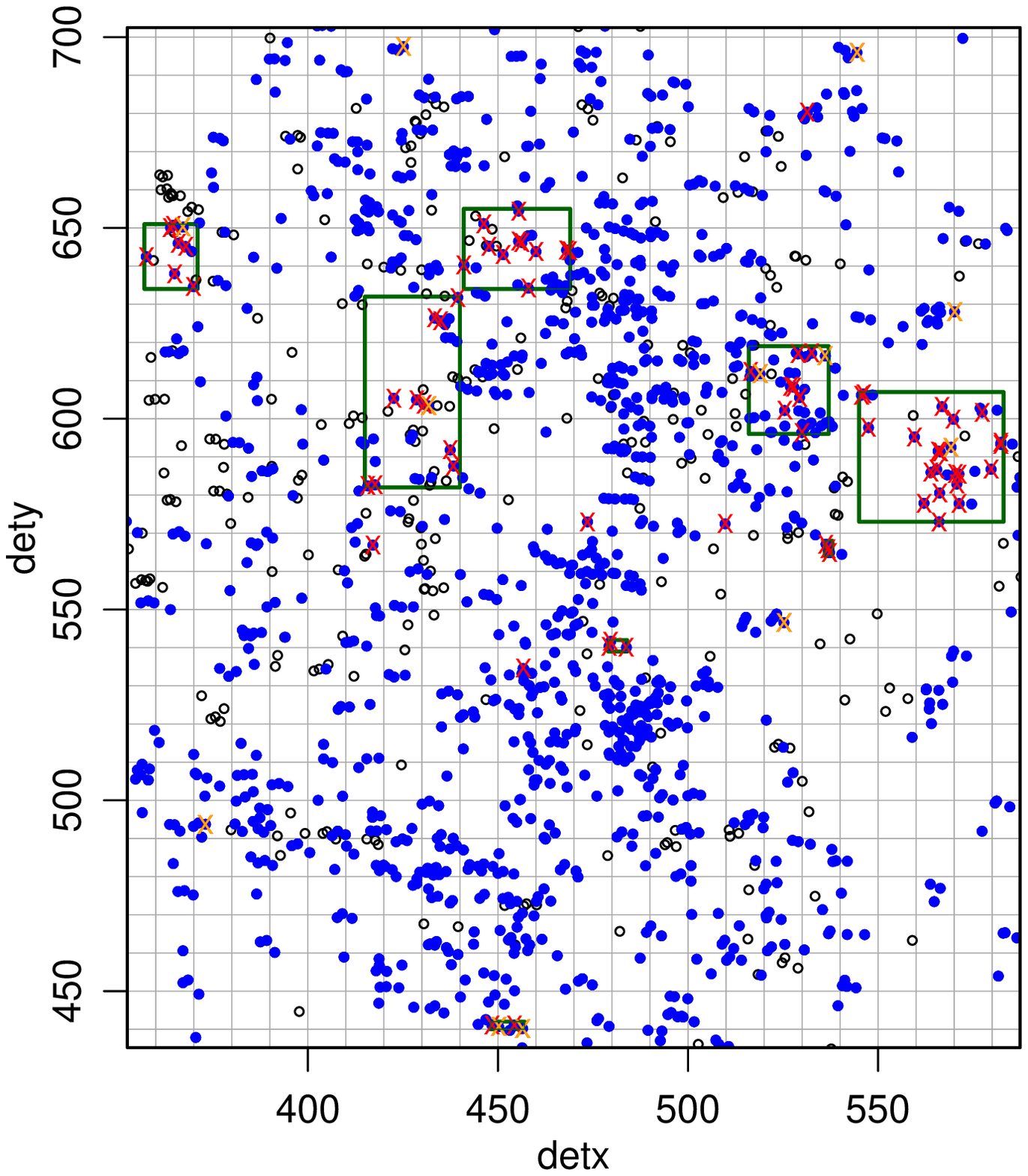} 
\caption{The detector region containing all the observed UVOT dropouts.
The black circles, blue dots and red Xs have the same meaning as in Figure~A2.
Green rectangles are drawn around clusters of three or more dropouts.  
Eight such rectangles enclose most red Xs and are used to define suspect regions on the UVOT detector.
All data points within these regions were eliminated to form the final light curve (Figure~2), regardless of whether or not they failed the dropout test.}
\label{fig:figa3}
\end{center}
\end{figure}

\end{document}